\documentclass[runningheads]{llncs}

\usepackage[T1]{fontenc}
\usepackage{amsmath,amssymb}
\usepackage{booktabs}
\usepackage{comment}
\usepackage{graphicx}
\usepackage{url}
\usepackage{multirow}
\usepackage{placeins}

\newif\ifanonymous
\anonymoustrue     % double-blind anonymous submission
\anonymousfalse  % camera-ready/non-anonymous version

\title{Loud-loss: A Perceptually Motivated Loss Function for Speech Enhancement Based on Equal-Loudness Contours}
\titlerunning{Loud-loss for Speech Enhancement}
\ifanonymous
\author{Anonymous Author(s)}
\authorrunning{Anonymous Author(s)}
\institute{Anonymous Affiliation(s)}
\else
\author{Zixuan Li\inst{1} \and
Xueliang Zhang\inst{1}\thanks{Corresponding author.} \and
Changjiang Zhao\inst{1} \and
Shuai Gao\inst{1} \and
Lei Miao\inst{2} \and
Zhipeng Yan\inst{2} \and
Sun Ying\inst{2} \and
Zhu Chong\inst{2}}
\authorrunning{Z. Li et al.}
\institute{College of Computer Science, Inner Mongolia University, Hohhot, China\\
\email{cslzx@mail.imu.edu.cn, cszxl@imu.edu.cn}
\and
Lenovo, Beijing, China}
\fi

\begin{document}

\maketitle

\begin{abstract}
    The mean squared error (MSE) is a ubiquitous loss function for speech enhancement, but its problem is that the error cannot reflect the auditory perception quality. This is because MSE causes models to over-emphasize low-frequency components which has high energy, leading to the inadequate modeling of perceptually important high-frequency information. To overcome this limitation, we propose a perceptually-weighted loss function grounded in psychoacoustic principles. Specifically, it leverages equal-loudness contours to assign frequency-dependent weights to the reconstruction error, thereby penalizing deviations in a way aligning with human auditory sensitivity. The proposed loss is model-agnostic and flexible, demonstrating strong generality. Experiments on the VoiceBank+DEMAND dataset show that replacing MSE with our loss in a GTCRN model elevates the WB-PESQ score from 2.17 to 2.93—a significant improvement in perceptual quality.

\end{abstract}

\section{Introduction}

Speech enhancement aims to improve speech quality and intelligibility by suppressing background noise, and serves as a key front-end module for downstream applications such as automatic speech recognition (ASR) \cite{weninger2015speech,zhang2017speech} and hearing aids \cite{wang2017deep,lesica2021harnessing}. Recently, deep learning-based methods have led to significant advances in speech enhancement, achieving remarkable improvements in both speech quality and intelligibility. Many methods operate in the spectral domain and learn to estimate a clean speech representation (e.g., magnitude spectra or complex masks) from noisy inputs \cite{tan2018convolutional,hu2020dccrn,li2025tf,fan2025bsdb}.

These models are commonly trained by minimizing the mean squared error (MSE) between estimated and clean spectral representations. Although MSE is attractive for its simplicity and stable optimization, it correlates imperfectly with human perception, leading to two key limitations. First, MSE assigns uniform penalties across frequency bins, regardless of perceptual importance. This is problematic because speech energy is highly uneven across frequency: vowels and voiced components usually dominate the low-frequency energy, whereas many consonant cues, harmonic details, and noise-coloration artifacts lie in regions with much lower absolute energy. Consequently, the optimizer can reduce a uniformly weighted squared error mainly by fitting energetic low-frequency components, while perceptually informative low-energy components remain blurred or under-estimated. Second, minimizing the spectral Euclidean distance, as MSE does, does not reliably predict or guarantee improvements in perceived speech quality \cite{10094773,manocha2020differentiable,fu21_interspeech}.

To address the uniform-penalty issue in MSE, several studies apply MSE to a compressed magnitude spectrum, such as $\mathcal{L}(\hat{S}, S) = MSE(\vert \hat{S} \vert^{\alpha}, \vert S \vert^{\alpha})$ for $\alpha \in (0,1)$ \cite{lee2018phase,shetu2024ultra}. This non-linear compression expands low-energy components and encourages more balanced error allocation across frequencies. However, it does so by changing the numerical scale of the error rather than by explicitly modeling where the ear is more sensitive. Two frequency bins with similar compressed magnitudes can still have different perceptual importance, so compression remains an amplitude-based remedy rather than a psychoacoustic weighting rule. Moreover, the choice of $\alpha$ is largely empirical without explicit psychoacoustic grounding, and can be sensitive to datasets and training conditions.

To better align training objectives with perception, other methods incorporate perceptually motivated targets. For example, Perceptual Contrast Stretching (PCS) \cite{chao2022perceptual} modifies the target magnitude spectrum using a power-law transformation weighted by a band importance function \cite{ANSI1997_S3.5}. While this can emphasize perceptually salient features in the target, it creates an input–target mismatch by training the network to map an original input to a distorted target representation.

In addition, several works introduce more explicit perceptual supervision for speech enhancement. A representative line uses learned speech representations as perceptual features, e.g., phone-fortified perceptual loss (PFPL) \cite{hsieh21_interspeech}, which measures Wasserstein distance \cite{olkin1982distance} between enhanced and clean signals in a pretrained wav2vec \cite{schneider2019wav2vec} latent space. Another line directly optimizes black-box perceptual metrics via adversarially learned surrogates, such as MetricGAN \cite{fu2019metricgan}. While effective, these approaches typically introduce auxiliary networks or depend on large pretrained models, increasing training complexity.

To address these limitations, we propose \textbf{Loud-loss}, a lightweight and interpretable training objective for speech enhancement. Loud-loss derives frequency weights from equal-loudness contours \cite{suzuki2004equal} and applies them to spectral reconstruction errors, so that errors in perceptually sensitive regions receive stronger penalties. Unlike feature-space or metric-learning objectives, it does not require auxiliary networks, pretrained representations, or learned metric predictors; it keeps the clean spectrum as the target and only changes how reconstruction errors are weighted. This design preserves input-target consistency while guiding capacity-limited models to allocate more attention to perceptually important frequency regions. We evaluate Loud-loss across architectures, output formulations, datasets, and perceptual baselines to verify its generality.
\ifanonymous
The code will be released upon acceptance.
\else
The code is available in camera-ready version.
% The code is available at \url{https://github.com/zx1292982431/Loud-Loss}.
\fi

\section{Method}

\begin{figure}[!ht]
\centering
\includegraphics[width=\textwidth]{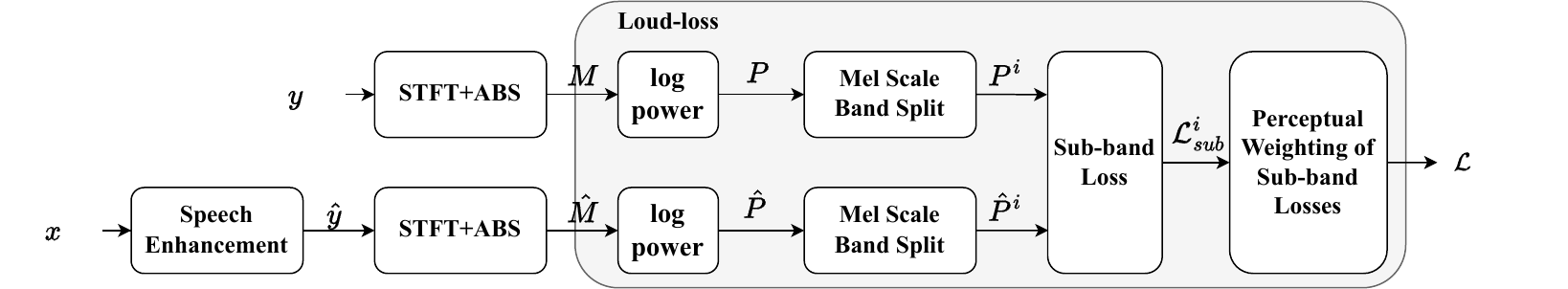}
\caption{Overview of the Loud-loss computation pipeline.}
\label{fig:pipeline} % Corrected typo
\end{figure}

As shown in Fig.~\ref{fig:pipeline}, Loud-loss is computed in four stages. First, the enhanced and target magnitude spectra are converted to the log-power domain. Second, the log-power spectra are partitioned into overlapping Mel-scaled sub-bands. Third, a loss is computed for each sub-band. Finally, these sub-band losses are weighted based on a psychoacoustic model and aggregated into a single loss value.

\subsection{Log-Power Spectrum Representation}
Loud-loss operates in the log-power spectral domain. Accordingly, the enhanced magnitude spectrum $\hat{M} \in \mathbb{R}^{F \times T}$ and the target magnitude spectrum $M \in \mathbb{R}^{F \times T}$ are first converted to their log-power representations, denoted as $\hat{P}$ and $P$. Here, $F$ and $T$ represent the number of frequency bins and time frames, respectively. This choice is motivated by two key factors: First, it aligns with the logarithmic nature of human sound intensity perception. Second, because equal-loudness contours are expressed in sound-pressure-level units, the log-power domain keeps the spectral reconstruction error compatible with the dB-domain perceptual weights described in Sec.~\ref{weight compute}. This avoids applying dB-derived weights directly to linear magnitude errors, where the error scale and the perceptual prior would be mismatched.

\subsection{Mel Scale Band Split}
\label{mel-band-split}

To mimic the non-uniform frequency resolution of the human auditory system, we partition the spectrum into $K$ overlapping sub-bands using the Mel scale. This process begins by defining $K+2$ boundary frequencies $\{f_{c}[i]\}_{i=0}^{K+1}$ that are equally spaced on the Mel scale. The conversion between frequency in Hertz (Hz) and the Mel scale is governed by \cite{o1987speech}:
\begin{equation}\label{eq:hz_to_mel}
mel = 2595 \cdot \log_{10}(1 + Hz / 700)
\end{equation}
\begin{equation}\label{eq:mel_to_hz}
Hz = 700 \cdot (10^{mel/2595} -1)
\end{equation}
These boundary frequencies are then mapped to their corresponding frequency bin indices $\{k_{c}[i]\}_{i=0}^{K+1}$. 
To obtain 50\% overlap, the $i$-th sub-band of a log-power spectrum $P$, denoted $P^{i} \in \mathbb{R}^{F_{i} \times T}$, is formed by grouping the frequency bins from index $k_{c}[i]$ to $k_{c}[i+2]-1$. This overlap is introduced to reduce discontinuities at sub-band boundaries, while the Mel spacing makes the grouping perceptually meaningful by providing higher resolution at lower frequencies and lower resolution at higher frequencies. The number of bins in the $i$-th sub-band is $F_{i} = k_{c}[i+2] - k_{c}[i]$. This process is applied to both $\hat{P}$ and $P$ to obtain the sub-band pairs $\hat{P^{i}}$ and $P^{i}$.

\subsection{Sub-band Loss Computation}
For each of the $K$ sub-bands, we compute the mean squared error (MSE) between the enhanced and target log-power spectra. The loss for the $i$-th sub-band, $\mathcal{L}_{sub}^{i}$, is defined as:
\begin{equation}\label{eq:sub_loss}
\mathcal{L}_{sub}^{i} = \frac{1}{F_i T} \sum_{f=1}^{F_i} \sum_{t=1}^{T} (P^i(f,t) - \hat{P}^i(f,t))^2
\end{equation}

The normalization by $F_iT$ makes each sub-band term an average error rather than a total error. This matters because Mel-spaced sub-bands contain different numbers of STFT bins after conversion to the discrete frequency grid. Without normalization, wider sub-bands would receive larger losses simply because they contain more bins, partially obscuring the intended perceptual weighting.

\subsection{Perceptual Weighting of Sub-band Losses}
\label{weight compute}
Our weighting scheme is designed to penalize errors in perceptually sensitive frequency regions more heavily. We derive perceptual importance weights from the 40-phon equal-loudness contour \cite{suzuki2004equal}, as this loudness level is representative of typical conversational speech. This curve shows that human hearing sensitivity varies with frequency; a lower Sound Pressure Level (SPL) required on the 40-phon contour at a given frequency indicates that less acoustic level is needed to achieve the same perceived loudness, implying higher auditory sensitivity in that region.

Since 1000 Hz is the specific reference frequency used to measure equal-loudness contours, we define the perceptual weight $w^{i}$ for each sub-band as the inverse of its 40-phon hearing threshold relative to this reference:
\begin{equation}\label{eq:weights_v1}
w^i = \frac{SPL(1000 \text{ Hz})}{SPL(f_{c}[i+1])}, \quad i = 0, 1, \dots, K-1
\end{equation}
where $f_{c}[i+1]$ is the center frequency of the $i$-th sub-band, and the $SPL(\cdot)$ function is implemented via a nearest-neighbor lookup on the values in Table \ref{tab:tab1_8col}. This formulation assigns a larger weight to frequency bands where human hearing is more sensitive (i.e., those with lower SPL thresholds).

\begin{table}[h!]
\centering
\caption{Required SPL (dB) versus frequency (Hz) for the 40-phon equal-loudness contour.}
\scriptsize
\setlength{\tabcolsep}{3.5pt}
\begin{tabular}{@{}rrrrrrrr@{}}
\toprule
\textbf{Freq} & \textbf{SPL} & \textbf{Freq} & \textbf{SPL} & \textbf{Freq} & \textbf{SPL} & \textbf{Freq} & \textbf{SPL} \\
\textbf{(Hz)} & \textbf{(dB)} & \textbf{(Hz)} & \textbf{(dB)} & \textbf{(Hz)} & \textbf{(dB)} & \textbf{(Hz)} & \textbf{(dB)} \\
\midrule
20 & 99.85 & 25 & 93.94 & 31.5 & 88.17 & 40 & 82.63 \\
50 & 77.78 & 63 & 73.08 & 80 & 68.48 & 100 & 64.37 \\
125 & 60.59 & 160 & 56.70 & 200 & 53.41 & 250 & 50.40 \\
315 & 47.58 & 400 & 44.98 & 500 & 43.05 & 630 & 41.34 \\
800 & 40.06 & 1000 & 40.01 & 1250 & 41.82 & 1600 & 42.51 \\
2000 & 39.23 & 2500 & 36.51 & 3150 & 35.61 & 4000 & 36.65 \\
5000 & 40.01 & 6300 & 45.83 & 8000 & 51.80 & 10000 & 54.28 \\
12500 & 51.49 & & & & & & \\
\bottomrule
\end{tabular}
\label{tab:tab1_8col}
\end{table}

The final loss is the sum of these perceptually weighted sub-band losses:
\begin{equation}\label{eq:final_loss}
\mathcal{L} = \sum_{i=0}^{K-1} w^i \mathcal{L}_{sub}^i
\end{equation}

\section{Experiment}
\subsection{Networks}
To assess Loud-loss under different model-capacity regimes, we evaluate it on two recent speech enhancement networks: the low-complexity GTCRN \cite{rong2024gtcrn} and the medium-complexity ICCRN \cite{liu2023iccrn}. This choice tests whether the proposed loss is effective beyond a single backbone, including a compact model where frequency-wise capacity allocation is especially important.

Because Loud-loss is defined on magnitude spectra, we use a magnitude-only prediction task for all compared losses: each network estimates the clean magnitude spectrum and reconstructs the waveform with the noisy phase. Keeping the prediction target and reconstruction procedure fixed makes performance differences more directly attributable to the training objective rather than to complex-spectrum estimation choices. In addition, we remove the ERB operation from the original GTCRN, which merges features above 2~kHz and biases the model toward low-frequency components, to avoid confounding the effect of Loud-loss with this architectural prior.

\subsection{Datasets and Metrics}
We conduct experiments on two 16~kHz evaluation settings: VoiceBank+\allowbreak DEMAND \cite{botinhao2016investigating} and DNS3 \cite{reddy2021interspeech}. VoiceBank+\allowbreak DEMAND is constructed from the VoiceBank corpus \cite{veaux2013voice}, which contains 30 native English speakers, with 28 used for training and 2 for testing; its official training set includes 11{,}572 utterances mixed with 10 noise types, including 8 from DEMAND \cite{thiemann2013diverse} and 2 synthetic noises. Because clean references are available, this test set supports reference-based metrics such as PESQ, STOI, ESTOI, SNR, and SI-SNR.

For DNS3, we construct a supervised training set of 360{,}000 paired 10-second noisy/clean utterances, while evaluation is performed on the DNS3 blind test set without clean references. We therefore use non-intrusive perceptual predictors, including P.808 MOS \cite{naderi2020}, DNSMOS P.835 scores for speech quality, background noise, and overall quality (SIG/BAK/OVRL) \cite{naderi21_interspeech}, and NISQA \cite{mittag2021nisqa}, which estimates overall MOS and quality dimensions related to noise, coloration, discontinuity, and loudness.

\subsection{Implementation Details}
All audio is processed using a Short-Time Fourier Transform (STFT) with a 512-sample (32ms) Hanning window and a 256-sample (16ms) hop. Following the procedure in Section \ref{mel-band-split}, the resulting spectrum is partitioned into $K=25$ Mel-scaled sub-bands, upon which the Loud-loss is computed.
The network input is formed by concatenating the real part, imaginary part, and magnitude of the noisy spectrogram along the channel axis. All models are trained using the Adam optimizer\cite{kingma2014adam} with an initial learning rate of 0.001 and a batch size of 4. We apply gradient clipping with a threshold of 5.0 to prevent exploding gradients. A learning rate scheduler is employed, which halves the learning rate if the validation loss fails to decrease for 20 consecutive epochs. Each model is trained for 200 epochs, and the checkpoint yielding the highest PESQ\cite{rix2001perceptual} score on the validation set is saved for testing.

Unless otherwise specified, the same STFT configuration and training schedule are used for all compared losses, since frame length, hop size, validation criterion, and optimizer settings can independently affect perceptual scores.

\subsection{Results on VCTK+Demand}
\subsubsection{Comparison with the baseline losses.}

\begin{table}[t]
    \centering
    \caption{Objective evaluation results on the VoiceBank+DEMAND test set. Boldface indicates the best score in each column. $^\dagger$ denotes magnitude mapping-based methods, while $^\ddagger$ denotes masking-based methods.}
    \scriptsize
    \setlength{\tabcolsep}{2.5pt}
    \renewcommand{\arraystretch}{1.08}
    \begin{tabular*}{\textwidth}{@{\extracolsep{\fill}}lcccccc@{}}
         \toprule
         \textbf{Exp.~ID} & \textbf{Models} & \textbf{Type} & \textbf{Loss} & \textbf{WB-PESQ} & \textbf{NB-PESQ} & \textbf{ESTOI} \\
         \midrule
         - & Unprocessed & - & - & 1.97 & 2.88 & 0.787 \\
         \midrule
         1 & GTCRN$^{\dagger}$ & Mapping & MSE & 2.17 & 3.04 & 0.812 \\
         2 & GTCRN$^{\dagger}$ + PCS & Mapping & MSE & 2.48 & 3.20 & 0.807 \\
         3 & GTCRN$^{\dagger}$ & Mapping & Loud-loss & \textbf{2.93} & 3.51 & 0.836 \\
         \addlinespace[2pt]
         4 & GTCRN$^{\ddagger}$ & Masking & MSE & 2.57 & 3.24 & 0.824 \\
         5 & GTCRN$^{\ddagger}$ & Masking & Loud-loss & 2.92 & 3.50 & 0.841 \\
         \addlinespace[2pt]
         6 & ICCRN$^{\dagger}$ & Mapping & MSE & 2.41 & 3.18 & 0.820 \\
         7 & ICCRN$^{\dagger}$ & Mapping & Loud-loss & 2.84 & 3.46 & 0.841 \\
         \addlinespace[2pt]
         8 & GTCRN$^{\dagger}$ & Mapping & Loud-loss + SI-SNR & 2.92 & \textbf{3.55} & \textbf{0.847} \\
         \addlinespace[2pt]
         9 & GTCRN$^{\dagger}$ & Mapping & Compressed$(\alpha = 0.3)$ & 2.85 & 3.47 & 0.844 \\
         % GTCRN$^{\dagger}$ & Mapping & Compressed$(\alpha = 0.5)$ & 2.82 & 3.44 & 0.839 & 0.939 & 18.43 & 17.07 \\
         10 & GTCRN$^{\dagger}$ & Mapping & Compressed$(\alpha = 0.7)$ & 2.64 & 3.36 & 0.832 \\
         \bottomrule
    \end{tabular*}

    \vspace{0.55em}

    \begin{tabular*}{\textwidth}{@{\extracolsep{\fill}}lcccccc@{}}
         \toprule
         \textbf{Exp.~ID} & \textbf{Models} & \textbf{Type} & \textbf{Loss} & \textbf{STOI} & \textbf{SNR} & \textbf{SI-SNR} \\
         \midrule
         - & Unprocessed & - & - & 0.921 & 8.54 & 8.44 \\
         \midrule
         1 & GTCRN$^{\dagger}$ & Mapping & MSE & 0.927 & 18.43 & 17.24 \\
         2 & GTCRN$^{\dagger}$ + PCS & Mapping & MSE & 0.924 & 14.20 & 8.67  \\
         3 & GTCRN$^{\dagger}$ & Mapping & Loud-loss & 0.936 & 16.28 & 14.41 \\
         \addlinespace[2pt]
         4 & GTCRN$^{\ddagger}$ & Masking & MSE & 0.930 & 19.30 & 17.96 \\
         5 & GTCRN$^{\ddagger}$ & Masking & Loud-loss & 0.937 & 18.78 & 17.43 \\
         \addlinespace[2pt]
         6 & ICCRN$^{\dagger}$ & Mapping & MSE & 0.920 & 18.43 & 17.23 \\
         7 & ICCRN$^{\dagger}$ & Mapping & Loud-loss & 0.935 & 17.81 & 16.53 \\
         \addlinespace[2pt]
         8 & GTCRN$^{\dagger}$ & Mapping & Loud-loss + SI-SNR & \textbf{0.940} & \textbf{20.37} & \textbf{18.44} \\
         \addlinespace[2pt]
         9 & GTCRN$^{\dagger}$ & Mapping & Compressed$(\alpha = 0.3)$ & \textbf{0.940} & 18.27 & 17.01 \\
         % GTCRN$^{\dagger}$ & Mapping & Compressed$(\alpha = 0.5)$ & 0.939 & 18.43 & 17.07 \\
         10 & GTCRN$^{\dagger}$ & Mapping & Compressed$(\alpha = 0.7)$ & 0.937 & 18.31 & 17.14 \\
         \bottomrule
    \end{tabular*}
    \label{tab:main_results}
\end{table}

\begin{figure}[!htbp]
    \centering
    \includegraphics[width=0.86\linewidth]{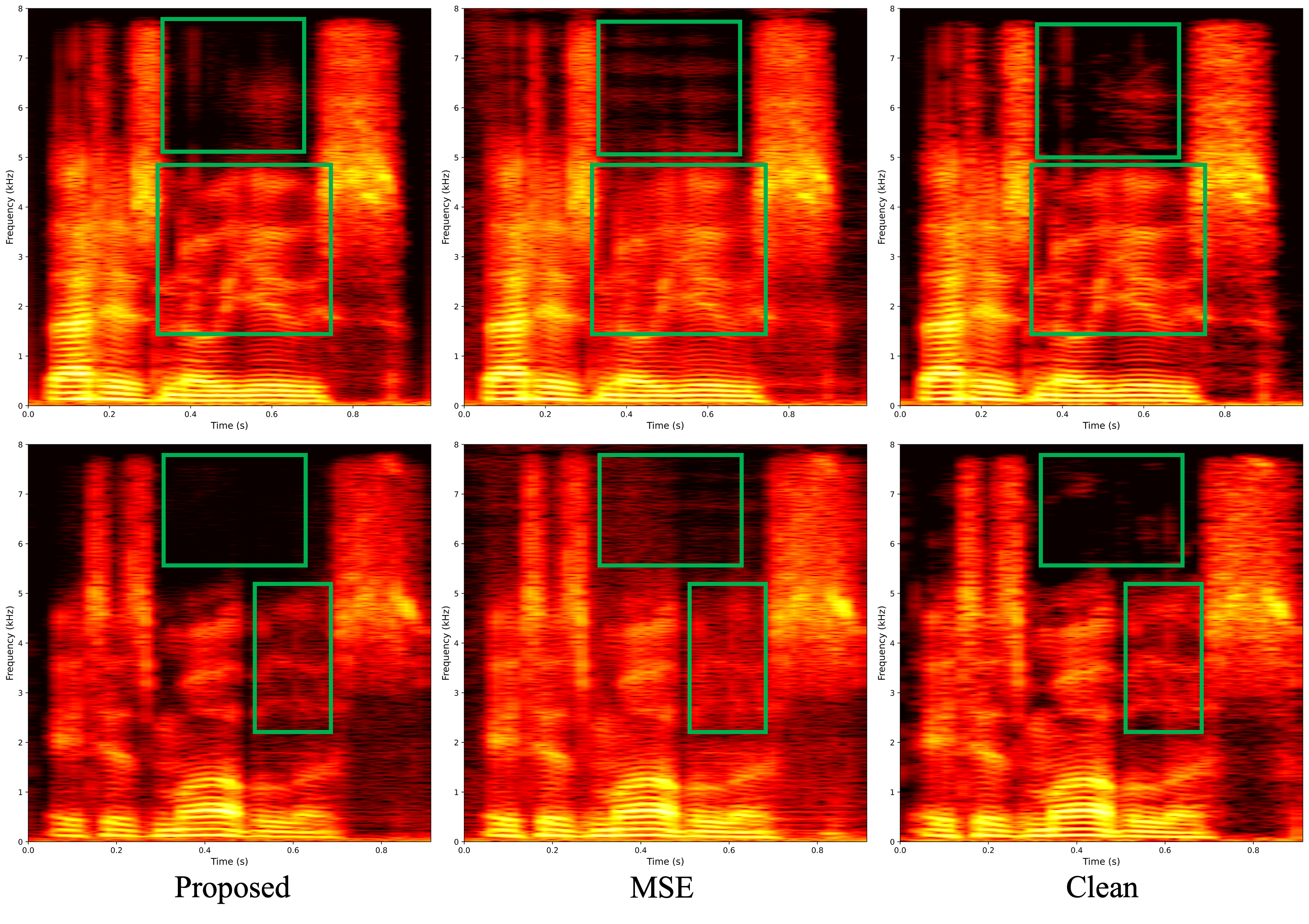}
    \caption{Comparison of enhanced speech spectrograms generated using different loss functions.}
    \label{fig:spectrogram_comparison}
\end{figure}

As shown in Table~\ref{tab:main_results}, the Loud-loss demonstrates significant advantages over the standard MSE baseline across all tested architectures. For the mapping-based GTCRN$^{\dagger}$ (Exp.~1 vs. 3), our method improves WB-PESQ by 0.76 and ESTOI by 0.024. This trend holds for the masking-based GTCRN$^{\ddagger}$ (Exp.~4 vs. 5) and the ICCRN model (Exp.~6 vs. 7). These substantial gains in perceptual metrics (PESQ) and intelligibility metrics (ESTOI) underscore the importance of integrating psychoacoustic principles into the loss function. The MSE loss, by uniformly weighting all frequency bins, forces the model to prioritize high-energy, low-frequency regions. This causes the blurring of higher-order harmonics and the inaccurate estimation of crucial high-frequency components, an effect visualized in the spectrogram analysis in Figure \ref{fig:spectrogram_comparison}. It is worth noting that while Loud-loss performs excellently on perceptual metrics, its scores on the SNR and SI-SNR metrics are lower than the MSE baseline. This is an expected trade-off. This trade-off is not a limitation but rather a validation of our approach. Metrics like SNR and SI-SNR are fundamentally aligned with the MSE objective—minimizing signal-level energy distance. The lower scores on these metrics, paired with superior PESQ and ESTOI scores, strongly indicate that Loud-loss successfully redirects the model's optimization focus from raw energy fidelity towards human perceptual fidelity, which is the central thesis of this work.

% \textbf{Generality and Flexibility:} The Loud-loss exhibits strong generality. First, it is effective for both mapping and masking-based enhancement frameworks, yielding substantial gains over MSE in both cases. Second, the loss is model-agnostic, providing consistent improvements when integrated into the architecturally distinct GTCRN and ICCRN models.

\textbf{Generality across Formulations and Models:} The comparison in Table~\ref{tab:main_results} indicates that the benefit of Loud-loss is not tied to a single output formulation or network architecture. For GTCRN, the gains appear in both the mapping setting, where the model directly estimates the clean magnitude spectrum, and the masking setting, where the model predicts a mask but is still guided by the perceptual error of the reconstructed magnitude. The improvement on ICCRN further shows that the same weighting strategy transfers to a distinct enhancement backbone. These results suggest that Loud-loss functions as a model-agnostic training criterion for spectral enhancement rather than as a method specialized to one architecture.

\textbf{Compatibility with Other Loss Functions:} The Loud-loss is also highly compatible with other losses. To demonstrate this, in Exp.~8, we combined it with an SI-SNR loss term. This composite loss achieved the best SI-SNR score of 18.44 (a substantial increase from 14.41 in Exp.~3) at the cost of only a negligible 0.01 decrease in WB-PESQ. This result highlights our method's flexibility, allowing for the creation of balanced, multi-objective loss functions that can be tailored to specific application requirements.

This compatibility is useful because different deployment scenarios may value different objective properties. A perceptual enhancement front end may prioritize PESQ and listening quality, whereas a communication or downstream-recognition system may also require lower signal distortion. Combining Loud-loss with SI-SNR retains the perceptual emphasis of equal-loudness weighting while recovering much of the signal-level accuracy favored by SI-SNR objectives.

\textbf{Comparison with Advanced Baselines:} Our method also outperforms other advanced training objectives, namely spectral compression (Exp.~9, 10) and the perceptually-motivated PCS (Exp.~2). Although spectral compression with an optimal $\alpha=0.3$ achieves competitive results by reducing the spectrum's dynamic range, our method still attains a higher WB-PESQ score (2.93 in Exp.~3 vs. 2.85 in Exp.~9). More importantly, our approach has a distinct advantage in interpretability: its weights are rigorously derived from a psychoacoustic model, whereas the optimal compression factor $\alpha$ must be found empirically and is highly sensitive, as shown by the performance drop when $\alpha$ is changed to 0.7. Furthermore, when compared to the perceptually-motivated PCS method (Exp.~2), which introduces an input-target mismatch, our method demonstrates comprehensive superiority, outperforming it across all reported metrics.

\subsubsection{Ablation studies.}

To validate the effectiveness of each design of Loud-loss, we conducted a series of ablation experiments using the mapping-based GTCRN$^{\dagger}$ as baseline. The results are shown in Table~\ref{tab:ablation}.

The ablations are organized around four design questions. We first examine the sub-band construction, including overlap, Mel spacing, and sub-band averaging. We then compare the equal-loudness weighting with simpler or alternative weighting rules. Next, we test whether applying the loss in the log-power domain is necessary. Finally, we study the interaction with GTCRN's ERB module to check whether the observed gains come from the proposed psychoacoustic weighting itself rather than from a particular implementation detail.

\begin{table}
    \centering
    \caption{Ablation study of the Loud-loss on the GTCRN$^{\dagger}$ model. "Baseline" refers to our full proposed method (Exp.~3).}
    \label{tab:ablation}
    \begin{tabular}{lcc}
        \toprule
        \textbf{Exp.~ID} & \textbf{Description} & \textbf{WB-PESQ}\\
        \midrule
        3 & Baseline (Loud-loss) & \textbf{2.93} \\
        \midrule
        \multicolumn{3}{l}{\textit{Analysis of Sub-band Splitting}} \\[-2pt]
        11 & w/o Sub-band Overlap & 2.81 \\
        12 & Uniform Band Split & 2.40 \\
        13 & Per-bin Weighting (no bands) & 2.56 \\
        \midrule
        \multicolumn{3}{l}{\textit{Analysis of Weighting Scheme}} \\[-2pt]
        14 & Uniform Weights & 2.60 \\
        15 & PCS-based Weights & 2.70 \\
        \midrule
        \multicolumn{3}{l}{\textit{Analysis of Application Domain}} \\[-2pt]
        16 & Applied on Magnitude & 2.05 \\
        \midrule
        \multicolumn{3}{l}{\textit{Analysis of ERB Interaction}} \\[-2pt]
        17 & Exp.1 with ERB & 2.30 \\
        18 & Exp.3 with ERB & 2.90 \\
        \bottomrule
    \end{tabular}
\end{table}

Our baseline (Exp.~3) employs a Mel-scale band split with 50\% overlap. First, we investigated the impact of this overlap. By removing it (Exp.~11), we observed a minor degradation in PESQ, which may be attributed to discontinuities at the sub-band edges during optimization. Next, we replaced the Mel scale with a uniform Hz scale for band splitting (Exp.~12). This change resulted in a substantial drop in WB-PESQ from 2.93 to 2.40, confirming that the perceptually-motivated Mel scale is crucial for allocating appropriate model capacity to different frequency regions. Finally, forgoing sub-bands entirely and instead applying interpolated weights directly to each frequency bin (Exp.~13) also yielded inferior results, suggesting that the sub-band averaging mechanism is beneficial.

To validate our proposed psychoacoustic weighting, we compare it against two alternatives. First, in Exp.~14, we applied a uniform weight to all sub-bands, making each contribute equally to the total loss. This approach, which is conceptually equivalent to an MSE loss in the log-power domain, led to a sharp performance drop, with the WB-PESQ score falling from 2.93 to 2.60. In Exp.~15, we replaced our equal-loudness-contour-derived weights with those from the PCS method, which are based on band importance function\cite{ANSI1997_S3.5}. It caused a significant decline in WB-PESQ (2.93 vs. 2.70). These results validate that our proposed weighting scheme, which is more directly tied to psychoacoustic models of loudness perception, is more effective at improving overall perceptual quality.

Applying the loss in the log-power domain is a critical design choice. We verify its importance in Exp.~16, where we applied our dB-derived perceptual importance weights directly to losses computed on the linear magnitude spectrum. This domain mismatch resulted in a severe performance degradation, with the WB-PESQ score plummeting from 2.93 to 2.05. This suggests that perceptual weighting is effective only when the loss is computed in a mathematically compatible domain (log-power), and that using dB-based weights on magnitude-domain errors weakens the intended perceptual emphasis.

We investigated the interaction between the loss function and the ERB module, which was removed from our main experiments. In Exp.~17, adding the ERB module to the GTCRN model trained with MSE (Exp.~1) improved the WB-PESQ score from 2.17 to 2.30. This shows that for a capacity-constrained model like GTCRN, using the ERB module to force the model to focus more on low-frequency components is a beneficial trade-off. In contrast, Exp.~18 shows that adding the ERB module to our proposed method causes a slight performance degradation, with WB-PESQ dropping from 2.93 to 2.90. This finding is particularly significant: it indicates that our loss function already provides a more effective and principled allocation of the model's limited capacity based on rigorous psychoacoustic principles, rendering the ERB module's heuristic not only redundant but slightly detrimental.

Taken together, the ablation results show that the effectiveness of Loud-loss comes from the combination of its main design choices rather than from frequency weighting alone. The loss works best when equal-loudness weights are applied to Mel-spaced sub-band errors in the log-power domain, while simpler alternatives such as uniform bands, per-bin weighting, uniform weights, or magnitude-domain application all reduce performance.

\subsection{Results on DNS3}
We compare Loud-loss with representative perceptual objectives (PFPL and MetricGAN) on DNS3. For fairness, all methods adopt the same enhancement backbone as Exp.~3 (GTCRN$^{\dagger}$), and only the training objective is changed.

\begin{table}[!htbp]
\centering
\caption{Evaluation on the DNS3 blind test set.}
\resizebox{\linewidth}{!}{
\begin{tabular}{lccccccccc}
\toprule
\multirow{2}{*}{\textbf{Loss}} 
& \multicolumn{4}{c}{\textbf{DNSMOS}} 
& \multicolumn{5}{c}{\textbf{NISQA}} \\
\cmidrule(lr){2-5} \cmidrule(lr){6-10}
& \textbf{P.808} & \textbf{SIG} & \textbf{BAK} & \textbf{OVRL}
& \textbf{MOS} & \textbf{Noise} & \textbf{Color} & \textbf{Disc.} & \textbf{Loud.} \\
\midrule
MSE        & 3.0477 & 2.9139 & 3.0105 & 2.3416 & 2.3468 & 2.0555 & 2.6022 & 3.0961 & 3.0108 \\
PFPL       & 3.2226 & \textbf{2.9757} & 3.4343 & 2.5276 & 2.4191 & 2.8315 & 2.7386 & 3.1316 & \textbf{3.1564} \\
MetricGAN  & 3.2266 & 2.9424 & 3.4958 & 2.5285 & \textbf{2.5918} & 2.9450 & \textbf{2.7597} & \textbf{3.5369} & 3.0876 \\
Loud-loss  & \textbf{3.2272} & 2.8972 & \textbf{3.6656} & \textbf{2.5500} & 2.5189 & \textbf{3.0390} & 2.7500 & 3.4316 & 3.1426 \\
\bottomrule
\end{tabular}
}
\label{tab:dns3_blind}
\end{table}

Table~\ref{tab:dns3_blind} reports the results on the DNS3 blind test set using two non-intrusive perceptual assessment tools, DNSMOS and NISQA. Compared with the MSE baseline, Loud-loss achieves consistent gains across both metric families, indicating that the proposed perceptual weighting improves generalization to real blind-test conditions. In DNSMOS, Loud-loss obtains the best P.808, BAK, and OVRL scores, with a clear gain in BAK, suggesting stronger suppression of residual background noise.
The NISQA results provide a more fine-grained view of this behavior. Loud-loss achieves the highest Noise score and a competitive Loudness score, which is consistent with its equal-loudness-contour-based design. Meanwhile, MetricGAN obtains the best NISQA MOS, Color, and Discontinuity scores, and PFPL slightly outperforms Loud-loss on the Loudness dimension. These results indicate that different perceptual objectives emphasize different aspects of speech quality. Overall, Loud-loss provides a favorable trade-off: it delivers the strongest DNSMOS overall quality and background-noise performance, while remaining competitive on NISQA without relying on auxiliary perceptual networks or large pretrained feature extractors.

\subsection{Training-Cost Comparison with Other Perceptually Motivated Methods}
Besides enhancement quality, training cost is an important practical factor when comparing perceptually motivated objectives. Table~\ref{tab:training_cost} reports a controlled benchmark on a 10-second utterance using the same GTCRN generator for all methods. This setting isolates the cost introduced by the training objective itself: MSE serves as the lightweight reconstruction baseline, PFPL represents feature-space supervision with a frozen pretrained wav2vec network, and MetricGAN represents metric-oriented adversarial training with an additional discriminator.
Table~\ref{tab:training_cost} uses a controlled single-GPU benchmark: all measurements use batch size 1, the same GTCRN generator, and one NVIDIA RTX 5090 GPU. Params (M) reports the training-time model size, and Memory (MB) reports peak GPU memory. Time (ms) reports the mean duration of one training step, averaged over 1000 measured steps after 100 warm-up steps; Rel. to MSE is the same time normalized by the MSE baseline. FLOPs (G) measures the computation of one training step, including one forward pass and one backward pass.

\begin{table}[!htbp]
\centering
\caption{Training-cost comparison with other perceptually motivated methods.}
\label{tab:training_cost}
\footnotesize
\setlength{\tabcolsep}{2.2pt}
\renewcommand{\arraystretch}{1.12}
\begin{tabular}{@{}c c c c c c c@{}}
\toprule
\textbf{Method} &
\textbf{Extra net} &
\shortstack{\textbf{Params}\\\textbf{(M)}} &
\shortstack{\textbf{Time}\\\textbf{(ms)}} &
\shortstack{\textbf{Rel.}\\\textbf{to MSE}} &
\shortstack{\textbf{Memory}\\\textbf{(MB)}} &
\shortstack{\textbf{FLOPs}\\\textbf{(G)}} \\
\midrule
MSE & None & 0.028 & $18.43{\pm}3.14$ & $1.00\times$ & 313.6 & 2.96 \\
PFPL & Frozen wav2vec & 32.565 & $56.65{\pm}4.88$ & $3.07\times$ & 871.1 & 296.14 \\
MetricGAN & Discriminator & 0.209 & $162.92{\pm}14.04$ & $8.84\times$ & 329.8 & 7.67 \\
Loud-loss & None & 0.028 & $19.38{\pm}2.88$ & $1.05\times$ & 313.6 & 2.96 \\
\bottomrule
\end{tabular}
% \begin{minipage}{0.98\linewidth}
% \footnotesize
%\textit{Note.} Step FLOPs are PyTorch-counted forward-and-backward FLOPs for one training step using \texttt{FlopCounterMode}. These counts are intended as a theoretical computational-cost reference and may undercount operations implemented outside standard PyTorch operators, such as STFT/iSTFT kernels or CPU PESQ computation. The wav2vec model is frozen during training, but its forward pass is required to compute the loss and backpropagate gradients to the generator.
% \end{minipage}
\end{table}

The results show that Loud-loss preserves nearly the same training footprint as MSE. It introduces no extra training-time network, keeps the same parameter count and GPU memory, and increases the measured step time by only $1.05\times$. In contrast, PFPL requires a frozen wav2vec forward path during training, increasing the active training-time parameters to 32.565M, the peak memory to 871.1 MB, and the measured step time to $3.07\times$ that of MSE. It also introduces substantial computational overhead, with one-step FLOPs increasing from 2.96G for MSE to 296.14G. MetricGAN has a smaller additional model than PFPL, but adversarial training requires the discriminator path and the corresponding metric-calculation steps, leading to the largest time overhead at $8.84\times$ the MSE step time. These comparisons indicate that Loud-loss provides perceptually motivated supervision while retaining the implementation simplicity and training efficiency of a reconstruction loss.

\FloatBarrier

\section{Conclusions}
We propose Loud-loss, a simple and interpretable perceptually motivated loss that uses equal-loudness contours to weight spectral errors according to human auditory sensitivity. Loud-loss improves perceptual quality and intelligibility on the VoiceBank+\allowbreak DEMAND dataset and maintains competitive non-intrusive perceptual performance on the DNS3 blind test set. The results also show that the method is compatible with complementary objectives such as SI-SNR, allowing a better balance between perceptual quality and signal-level accuracy. Since Loud-loss modifies only the training objective and leaves the inference model unchanged, it can be integrated into existing spectral enhancement systems with minimal architectural modification. Future work may explore other loudness levels, adaptive contour selection, or task-specific combinations with intelligibility-oriented losses.

\begingroup
\small
\subsubsection{Acknowledgments}
This work was supported by the Inner Mongolia Natural Science Foundation (2025LHMS06005), the Hohhot R\&D Investment Incentive Program (2025-Gui-Yan-1), the Inner Mongolia Postgraduate Research Project (KC2025006S), and the CCF-Lenovo Research Fund (No. 20240203).
\endgroup

\bibliographystyle{splncs04_unsrt}
\bibliography{mybib}

\end{document}